\documentclass[]{raa}            
\usepackage{graphicx,times}
\usepackage{natbib}

\providecommand{\bm}[1]{\mbox{\boldmath$#1$\unboldmath}}
\def\kms{$\rm{km~s}^{-1}$}

\def\coTwo{$^{12}$CO ($J$=2-1)\space}
\def\coThr{$^{12}$CO ($J$=3-2)\space}
\def\TcoTwo{$^{13}$CO ($J$=2-1)\space}
\def\TcoThr{$^{13}$CO ($J$=3-2)\space}
\def\NcoTwo{CO ($J$=2-1)\space}
\def\NcoThr{CO ($J$=3-2)\space}
\def\coOne{$^{12}$CO ($J$=1-0)\space}
\def\coFou{$^{12}$CO ($J$=4-3)\space}

\begin{document}

   \title{Radiation-driven Implosion in the Cepheus B Molecular Cloud
}

 \volnopage{ {\bf 2010} Vol.\ {\bf X} No. {\bf XX}, 000--000}
   \setcounter{page}{1}

   \author{Sheng Chen
      \inst{1,2}
   \and Maohai Huang
      \inst{1}
   }

   \institute{National Astronomical Observatories, Chinese Academy of Sciences,
             Beijing 100012, China; {\it mhuang@nao.cas.cn}\\
   \and {Graduate University of the Chinese Academy of Sciences}
        \vs \no \\
        {\small  Received 2010 Mar. 23; accepted 2010 Apr. 28  }
  }

\abstract{ We analyze large scale mapping observations of the
molecular lines in the \coTwo, \coThr, \TcoTwo, and  \TcoThr transition emissions
toward the Cepheus B molecular cloud with the KOSMA 3m-telescope.
The integrated intensity map of the \coTwo transition
has shown a structure with a compact core and a compact ridge
extended in the north-west of the core. The cloud is surrounded by an optically bright rim,
where the radiation-driven implosion (RDI) may greatly change the gas properties.
The intensities of the \NcoThr transition are higher than those of the \NcoTwo transition along
the rim area. We find characteristic RDI structure in position-velocity diagrams.
Non-LTE Large velocity gradient (LVG) model analysis shows that the
density and temperature at the edge are higher than that in the center.
Our results provide evidences that RDI is taking place in Cepheus B molecular cloud.
 }

   \authorrunning{S. Chen, M. Huang }            
   \titlerunning{Radiation-driven Implosion in the Cepheus B Molecular Cloud }  
   \maketitle


%
%
\section{Introduction}           
\label{sect:intro}

The association Cepheus OB3 is one of the youngest groups of early
type stars associated with the HII region S155
\citep{1959Sharpless} at a distance of 0.7 kpc \citep{1964Blaauw}. The
observations in the \coOne line have shown a few bright
components named Cepheus A, B, C, D, E, F (Sargent 1977, 1979). Of
the six components, the hottest CO feature and abundant dust
emission were detected in the Cepheus B, which was surrounded by an
arc-like optical HII region S155. The molecular cloud/HII region
complex with such a morphology is called bright rimmed cloud
(BRC), where star formation activity can be triggered by the
radiation-driven implosion (RDI) mechanism.

The radio continuum observations \citep{1978Felli} by Westerbork
Synthesis Radio Telescope (WSRT) confirmed that the UV radiation
from a bright O7 star created an arc-shaped ionization front which
surrounded the Cepheus B molecular cloud. The observations of the
H$_{2}$CO and recombination lines \citep{1981Panagia} have
suggested that the ionized material was flowing away at about 11
\kms, and the ionization front was moving into the molecular cloud
at a velocity of about 2 \kms. \coThr and \TcoTwo observations
together with the far-infrared analysis showed that the cloud was
externally heated and the edge of the cloud was compressed by the
 expansion of S155 \citep{1992Minchin}.

\citet{2000Beuther} observed the Cepheus B area at a large scale
in the $J$=3-2 and 2-1 transitions of $^{12}$CO, $^{13}$CO and
C$^{18}$O with the KOSMA 3m-telescope. Based on the volume
densities derived from PDR model, they found that Cepheus B was
highly clumped and the volume of these clumps fills only 2-4\% of
the whole cloud. \citet{2006Mookerjea} confirmed this with further
observations of [C\,I] $^3$P$_2$ - $^3$P$_1$ and \coFou. Based on
the studies of protoplanetary disks, \citet{2009Getman} found a
spatial-temporal gradient of young stars from the molecular core
 toward HD217086, which identified HD217086 as the primary
 ionizing source of the cloud.

Although a lot of efforts have been made to characterize the
physical conditions of the Cepheus B, the mechanism of the star
formation in this region is still not quite clear. We present
analysis of a larger scale observation of multiple CO lines, and discuss
the RDI as the possible trigger of the star formation in this region.

\section{Observations}
\label{sect:Obs}

The observations were carried out with the 3m KOSMA sub-millimeter
telescope using the On-The-Fly mode in March, 2004. Mapping
observations of CO were made at \coTwo (230.538 GHz), \coThr
 (345.789 GHz), \TcoTwo (220.399 GHz), and \TcoThr (330.588 GHz) transitions.
The dual-channel 230/345 GHz SIS receivers (Graf et al. 1998) were
used to simultaneously observe the two transitions of the
isotopes. The system temperatures were about 160 K for $J=2-1$ transitions and 300K for $J=3-2$ transitions. The
pointing accuracy was better than 15$''$ and the angular
resolutions (FWHM of beam) were 130$''$ at 220 GHz and 80$''$ at
330 GHz. The efficiency of the telescope F$_{\rm eff}$ was 93\%,
and the main beam efficiencies of B$_{\rm eff}$ were 68\% at 230GHz
and 72\% at 345GHz. Line intensities were converted on the main beam scale,
using $T_{\rm mb}=(F_{\rm eff}/B_{\rm eff})T_{A}^{*}$ \citep{1989Downes}.

For the \coTwo and \coThr transition emissions, a 15$\times$15 map were observed
centering at RA=\hms{22}{55}{20.708}, DEC=\dms{62}{20}{00.02}
(B1950), while for \TcoTwo and \TcoThr, a 12$\times$11-point
 map were made centering at RA=\hms{22}{55}{24.000},
DEC=\dms{62}{20}{41.01} (B1950). The grid spacing was 1$'$. Each
point of the grid was observed 5-6 times with a total integration
time of 2 minutes and the spectra were averaged in order to
increase the signal to noise ratio. The observed data were reduced
and plotted with CLASS and GREG of Gildas package (Guilloteau \&
Lucas, 2000). The \NcoThr transitions were smoothed to a
resolution of 130$''$ of \NcoTwo transitions.

\section{Results}
\label{sect:results} 
\begin{figure}

\includegraphics[width=52mm, angle=-90]{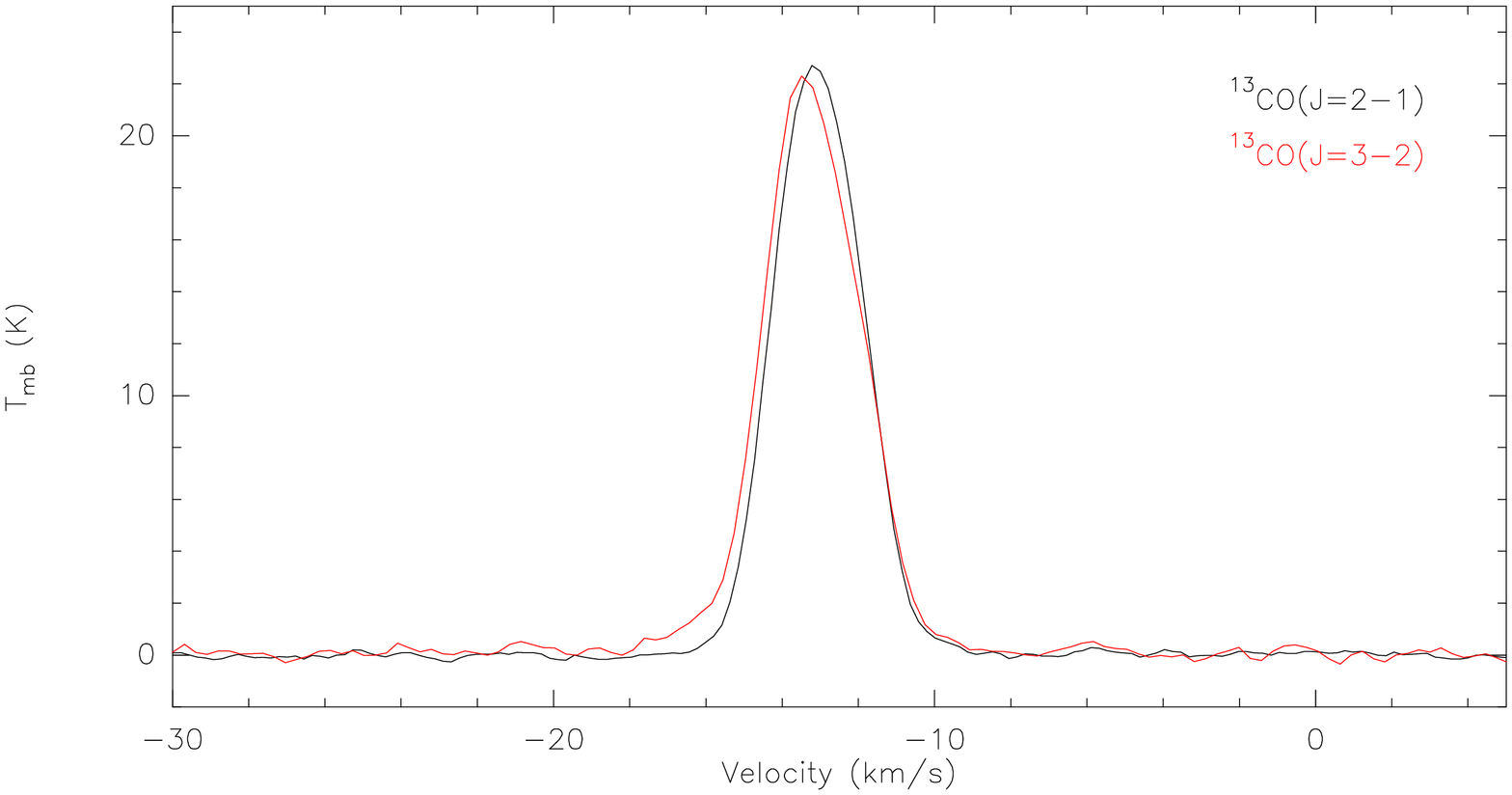}
\includegraphics[width=52mm, angle=-90]{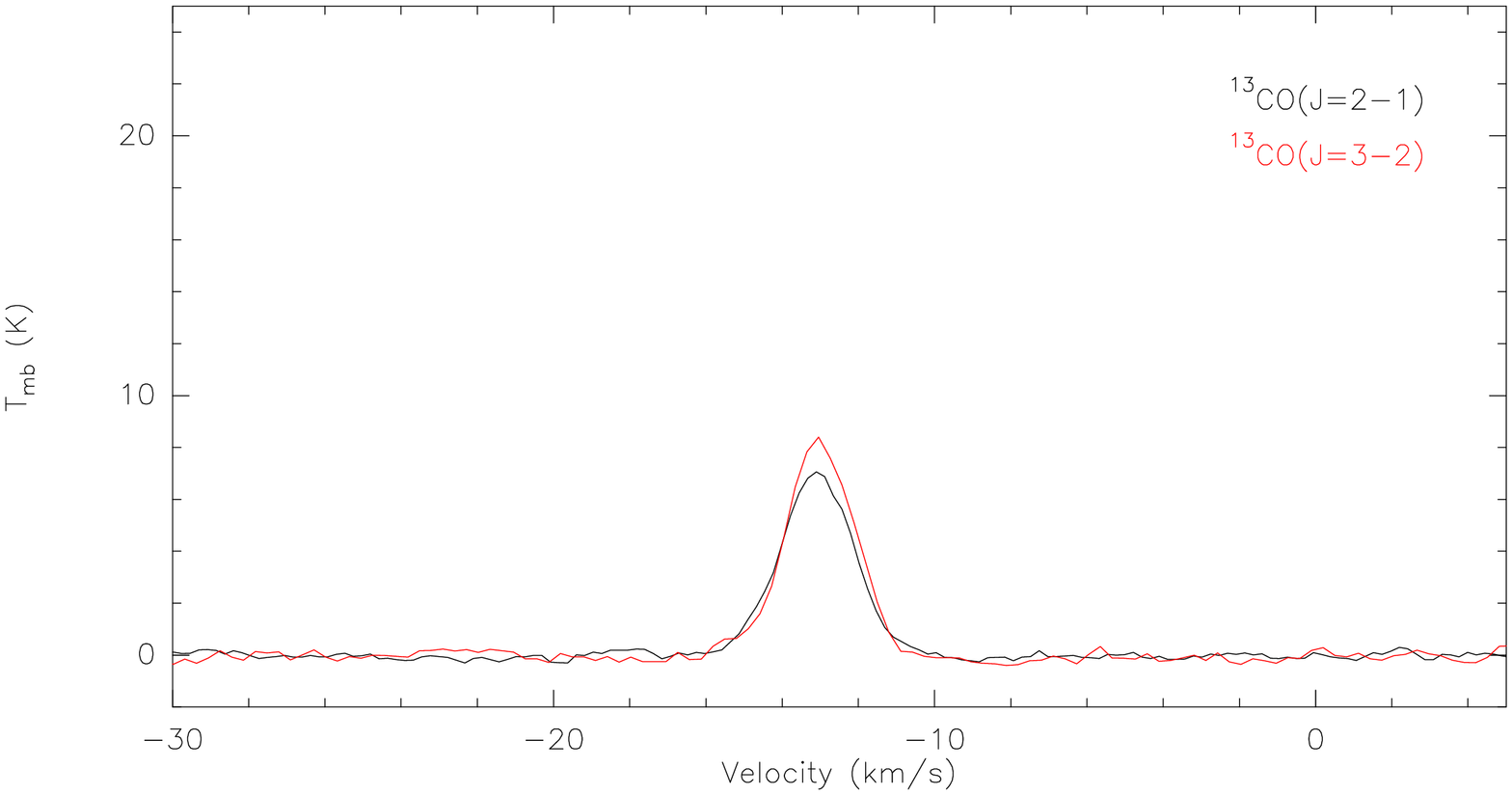}
\caption{Spectra at the center positions. The vertical axis is
main beam temperature T$_{\rm mb}$, and the horizontal axis is the
velocity.  Left panel: $^{12}$CO ($J$=2-1 and $J$=3-2), right panel:
$^{13}$CO ($J$=2-1 and $J$=3-2). (Black for $J$=2-1 transitions, and red
for $J$=3-2 transitions)}
   \label{Fig1}
\end{figure}

The spectra of \coTwo, \coThr, \TcoTwo, and
\TcoThr at the center positions of the observations
are shown in Figure \ref{Fig1}. The observed spectra show close to Gaussian line shapes.
From the 4 lines, the average v$_{\rm
LSR}$ is  $-13.1\pm0.1$ \kms.

\begin{figure}
   \centering
   \includegraphics[width=13.0cm, angle=0]{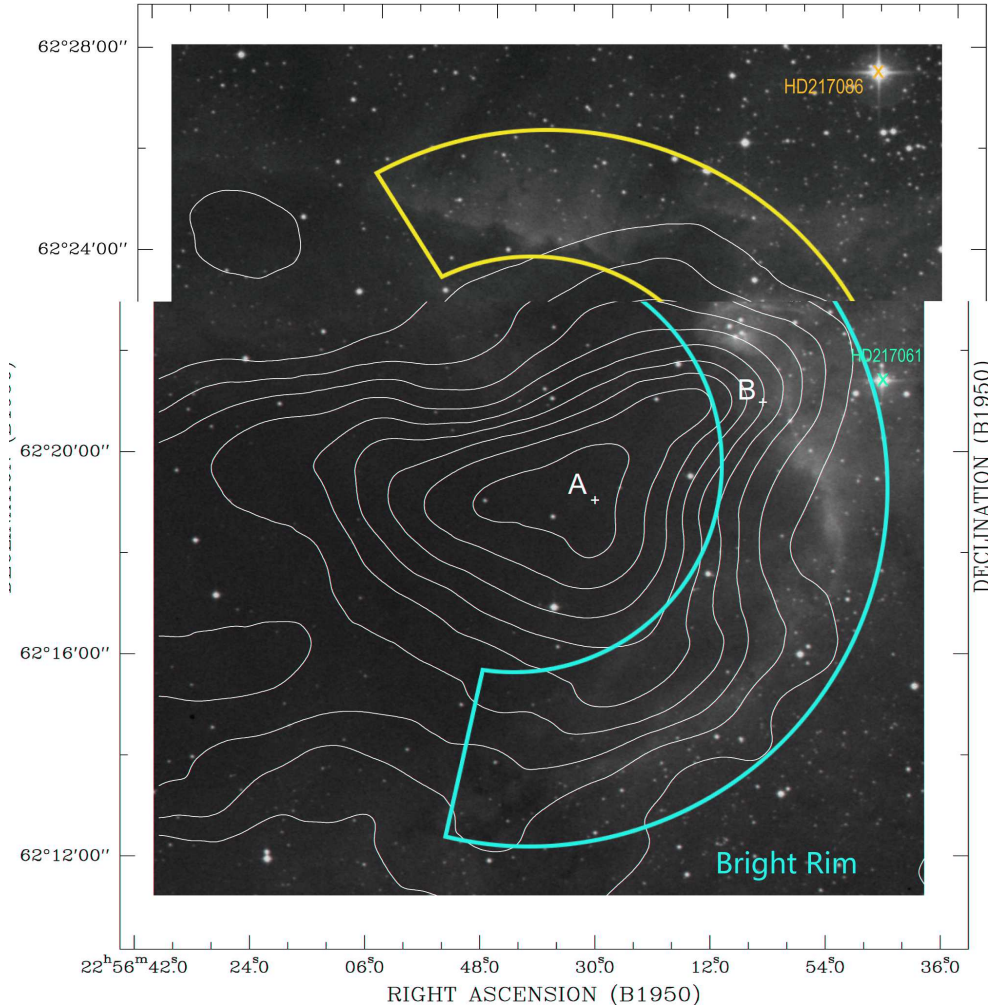}


\caption{ Contours of the \coTwo line integrated intensity from -18 to -8 \kms (white solid line)
overlaid on POSS2 Red image (gray scale). The contour levels of
solid lines start from 11 to 70 by step of 7.4 K\kms. Points A
and B are marked as typical positions at the center (A) and the
edge (B) of the cloud for further analysis. The bright rim in
optical image is marked in yellow.
 HD217086 and HD217061 are marked in orange cross.}
   \label{Fig2}
   \end{figure}

\begin{figure}
   \centering
   \includegraphics[width=8.0cm, angle=-90]{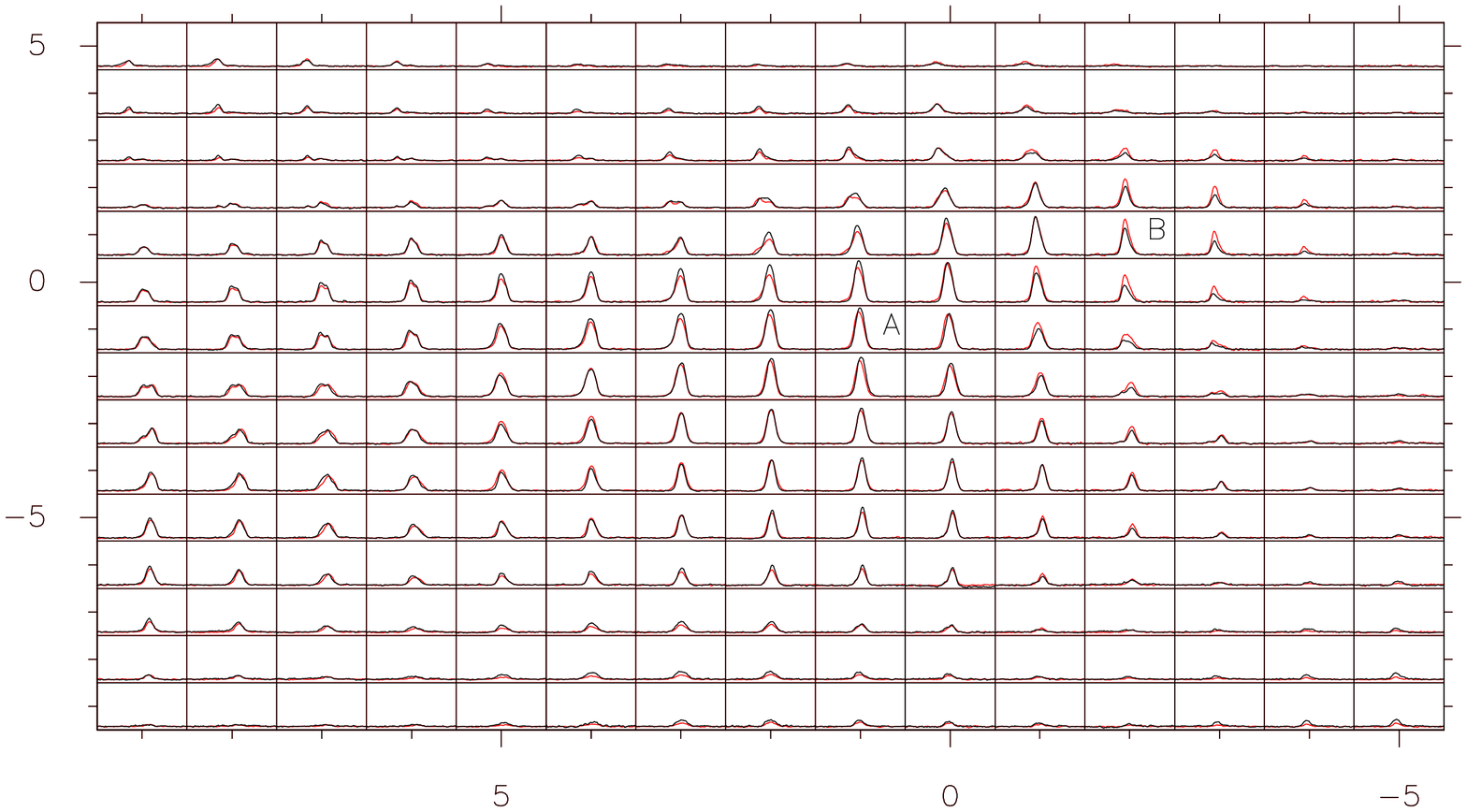}

\caption{ Spectra of the \coTwo (black) and \coThr (red) of all
grid points. For each grid point, the horizontal axis is the
velocity with a scale from -30 to 5\kms, and the vertical axis is
the main beam temperature from -2 to 25 K. A and B are the same
positions as marked in Figure \ref{Fig2}.}

   \label{Fig3}
   \end{figure}

We have made integration intensity maps of the \coTwo
line integrated from -18 to -8 \kms superimposed on the optical
POSS2 red plate as shown in Figure \ref{Fig2}.
O7 star HD217086 and B1 star HD217061 of Cepheus OB3 association are marked,
and the former is thought to be the primary ionizing source of the cloud \citep{2009Getman}.
The map has shown a
compact core and a compact ridge with a cometary shaped tail
extended in south-east direction. There is a steep drop-off in CO emission
at northern and western side of the cloud, forming a well-defined boundary edge.
Meanwhile an optically bright rim is shown
in north-west direction as seen in the POSS2 red plate in
gray scale. The compact ridge of the CO integrated intensity
contours is surrounded by the bright rim.
The average radial velocity of the Cepheus OB3 association is $-14.5\pm4.5$ \kms
\citep[][derived from optial spectra]{1973Garmany}, while in the north-western part the v$_{\rm LSR}$ of our CO observation is also about -15 to -14 \kms (see Figure \ref{Fig3}). The agreement between optical and CO velocities confirms that the bright rim and the molecular cloud are physically associated. The morphology of the
cloud and the HII region matches well with the scenario of
bright-rimmed clouds, which is possibly shaped by the
radiation-driven implosion of the ionizing source \citep{1989Bertoldi}.

In order to investigate the kinetic and physical conditions in
this region, we take position A (offset 1, -1) at the center of
 the cloud and position B (offset -2, 1) at the edge, as
marked in Figure \ref{Fig2}. The line properties are compared in
Table \ref{Table 1}. At position B, the \coThr transition intensity is higher than that of \coTwo, while at
position A, the \coTwo transition intensity is higher than that of \coThr.
The reversed line ratio suggests higher density or higher temperature at position B.
Further, we plot spectra of the \coThr (red) over \coTwo
(black) of all grid points in Figure \ref{Fig3}. We can see that the A and B positions
 are not alone in this aspect. The \coThr intensity is higher than \coTwo intensity
across the bright rim area for most grid positions, indicating the UV radiation is interacting
with the molecular cloud and pumping molecular to higher excitation levels.

\begin{table}[h!!!]

\small
\centering

\begin{minipage}[]{120mm}
\caption[]{ $^{12}$CO Line Parameters}
\label{Table 1}\end{minipage}

\tabcolsep 6mm
 \begin{tabular}{clcl}
  \hline\noalign{\smallskip}
Line    &v$_{\rm LSR}$(\kms)    &FWHM(\kms) &T$_{\rm mb}$(K)              \\
  \hline\noalign{\smallskip}
\coTwo (A)  &-12.6 (0.1)    &2.8 (0.1)  &24.2 (0.3)  \\
\coThr (A)  &-12.9 (0.1)    &3.0 (0.1)  &21.9 (0.4)  \\
\coTwo (B)  &-13.7 (0.1)    &2.4 (0.1)  &14.8 (0.8)  \\
\coThr (B)  &-13.6 (0.1)    &2.6 (0.1)  &19.7 (1.0)  \\
  \noalign{\smallskip}\hline
\end{tabular}

\end{table}

\begin{figure}
   \centering
   \includegraphics[width=5.0cm, angle=-90]{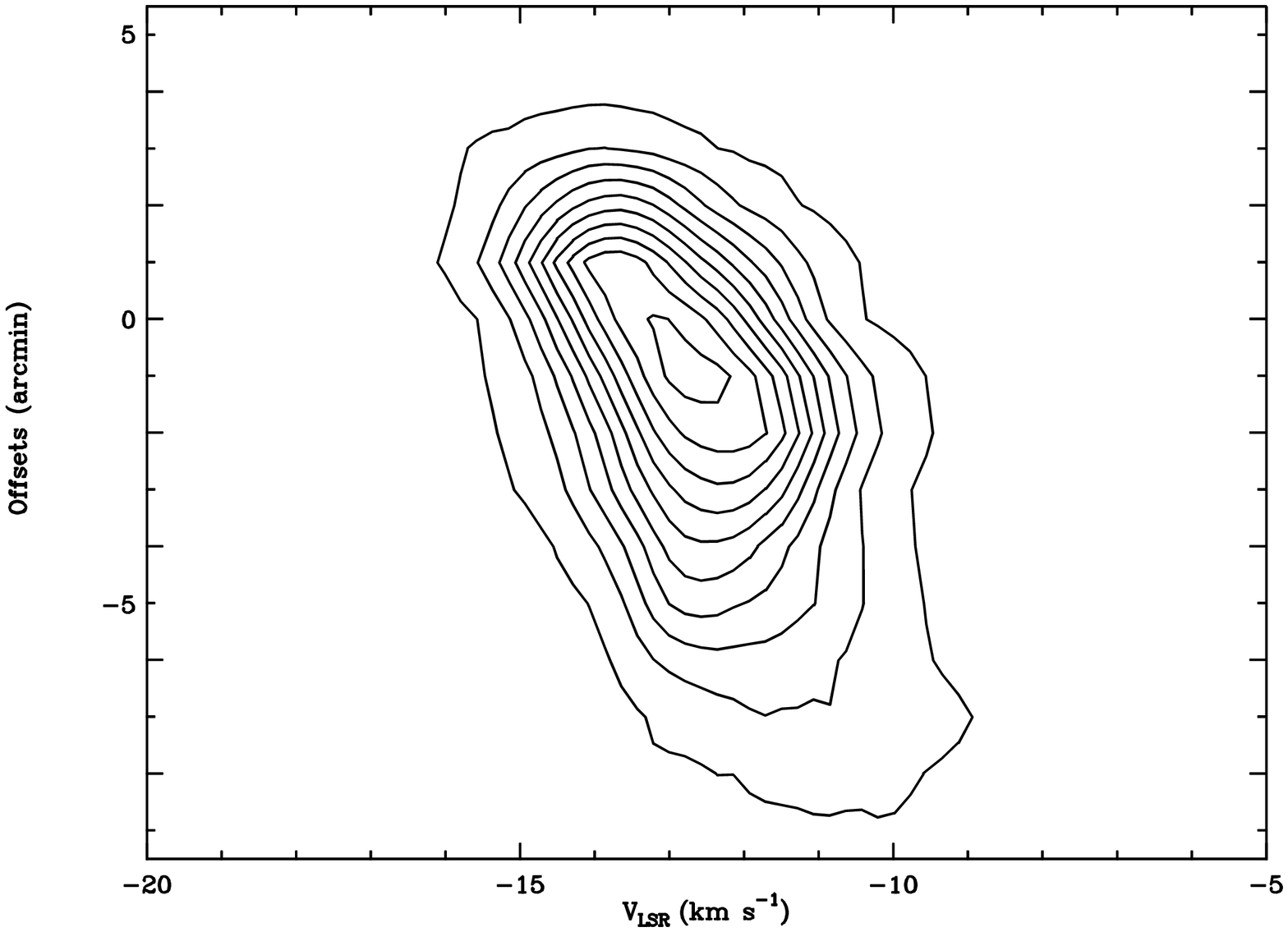}
   \includegraphics[width=5.0cm, angle=-90]{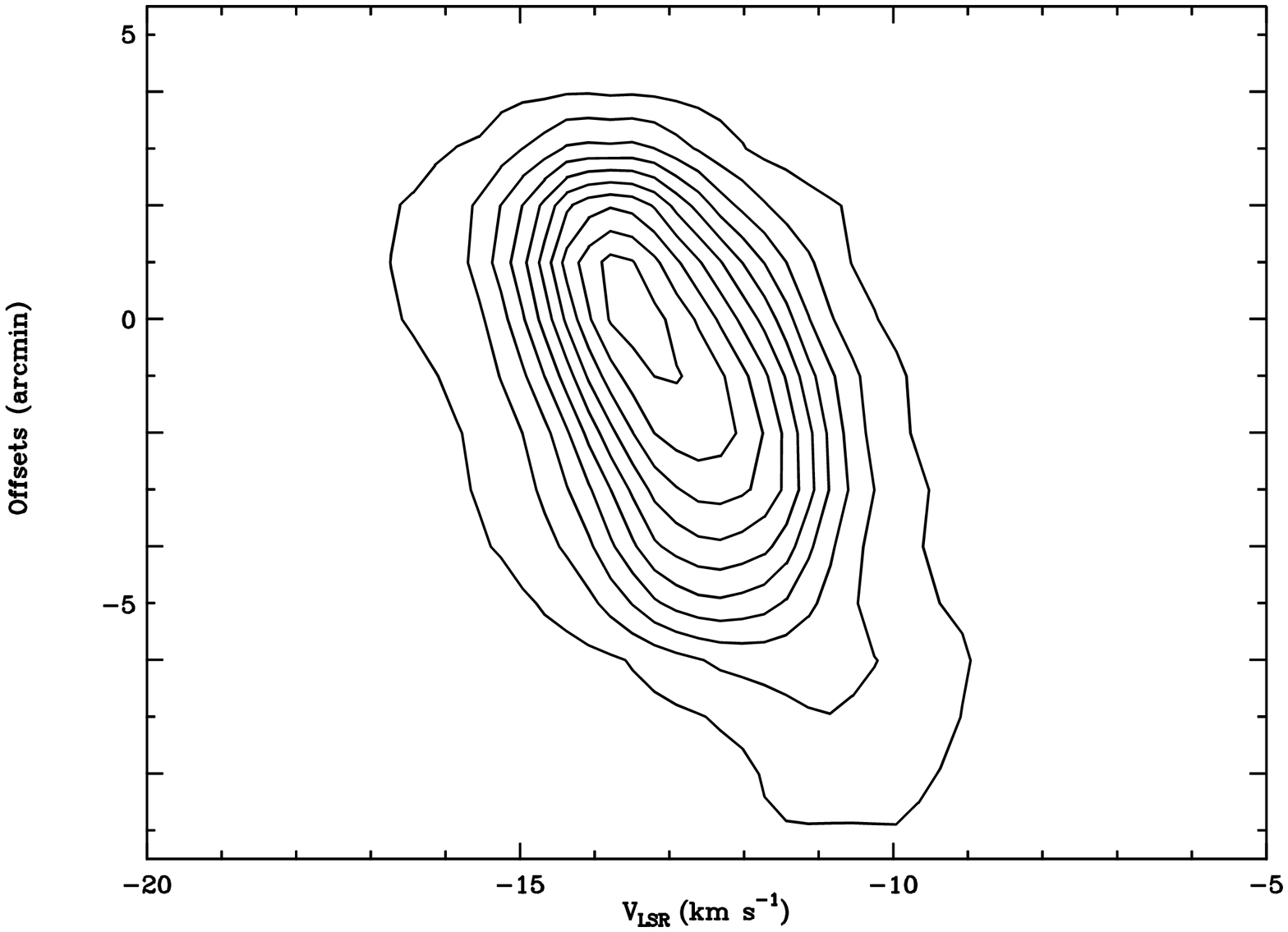}

\caption{ P-V diagram constructed from the \coTwo (left panel) and \coThr (right panel) along
the diagonal direction from north-west (positive offsets) to south-east (negative offsets).
}

   \label{Fig4}
   \end{figure}

From Figure \ref{Fig3} we also find a spatial shift of v$_{\rm LSR}$
from the head (north-west) to the tail (south-east) of the cloud,
which is clearly shown in position-velocity (PV) diagrams along the
diagonal direction from north-west to south-east in Figure \ref{Fig4}.

\section{Discussion}
\label{sect:discussion}

As seen in Figure \ref{Fig2}, the observed molecular cloud is
surrounded by an ionized bright rim facing the exciting source
with a hot core close to the rim and a cometary shaped tail on
the side opposite to the rim. This scenario matches well with
BRCs which is modeled as externally
illuminated, photo-evaporated and ablated into elongated head-tail
morphologies by UV radiation of OB stars \citep{1983Reipurth}. The
isolation and simple morphology of BRCs make them ideal
environments to investigate the RDI \citep{1989Bertoldi} mode of
star formation, where pressure from the ionization shock front at
the surface propagates through a cloud and overcomes the
magnetic, turbulent and thermal pressure trigger cloud collapse,
thereby triggering local star formation.

In our case, the spatial shift of v$_{\rm LSR}$ shows strong connections with the RDI
mechanism.
The PV diagrams in Figure \ref{Fig4}
shows a velocity gradient from the head (north-west) of the cloud to the tail (south-east).
The blueward elongation toward the head by about 3 \kms from the center indicates that the head part is expanding towards us,
while the redward elongation toward the tail by about 4 \kms indicates the tail part is flowing away to the opposite direction.
The velocity gradients of the head and tail matches with the RDI model prediction of \citet{1994Lefloch},
where the red-shift of the tail is caused by the acceleration of the gas across the tail, and the blue-shift of the head is because the compressed gas in the head moves faster than the gas immediately behind it. Such PV diagrams are reported by various BRC and cometary globule studies \citep{1995Lefloch, 1997Sugitani, 1997White, 2002Bachiller} which are explained by the RDI mechanism, and so are taken as a characteristic structure of the RDI.

The reversed line ratios across the bright rim are another interesting feature.
The spatial distribution of the reversed line ratios suggest that they are physically connected with the RDI along the rim area.
Higher transitions are more readily excited by higher temperatures or higher densities (Petitpas, 1998),
either of which can be a result of RDI. The highly luminous ionizing shock front from the RDI
interacts with the molecular cloud, pumping more particles from a lower
level to a higher one, which makes the $J$=3-2 emission stronger along the rim area.

\begin{figure}
\includegraphics[width=75mm, angle=0]{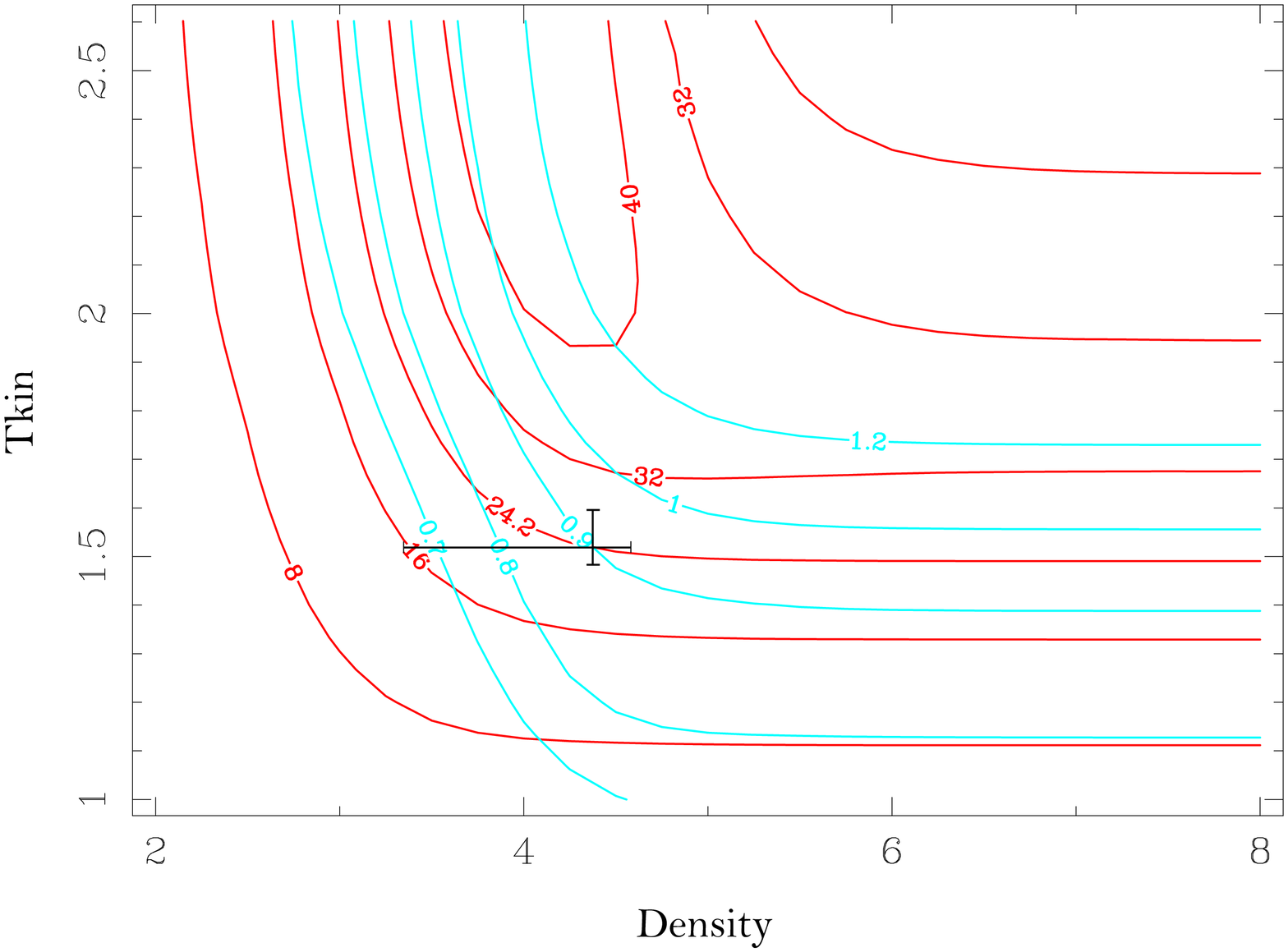}
\includegraphics[width=75mm, angle=0]{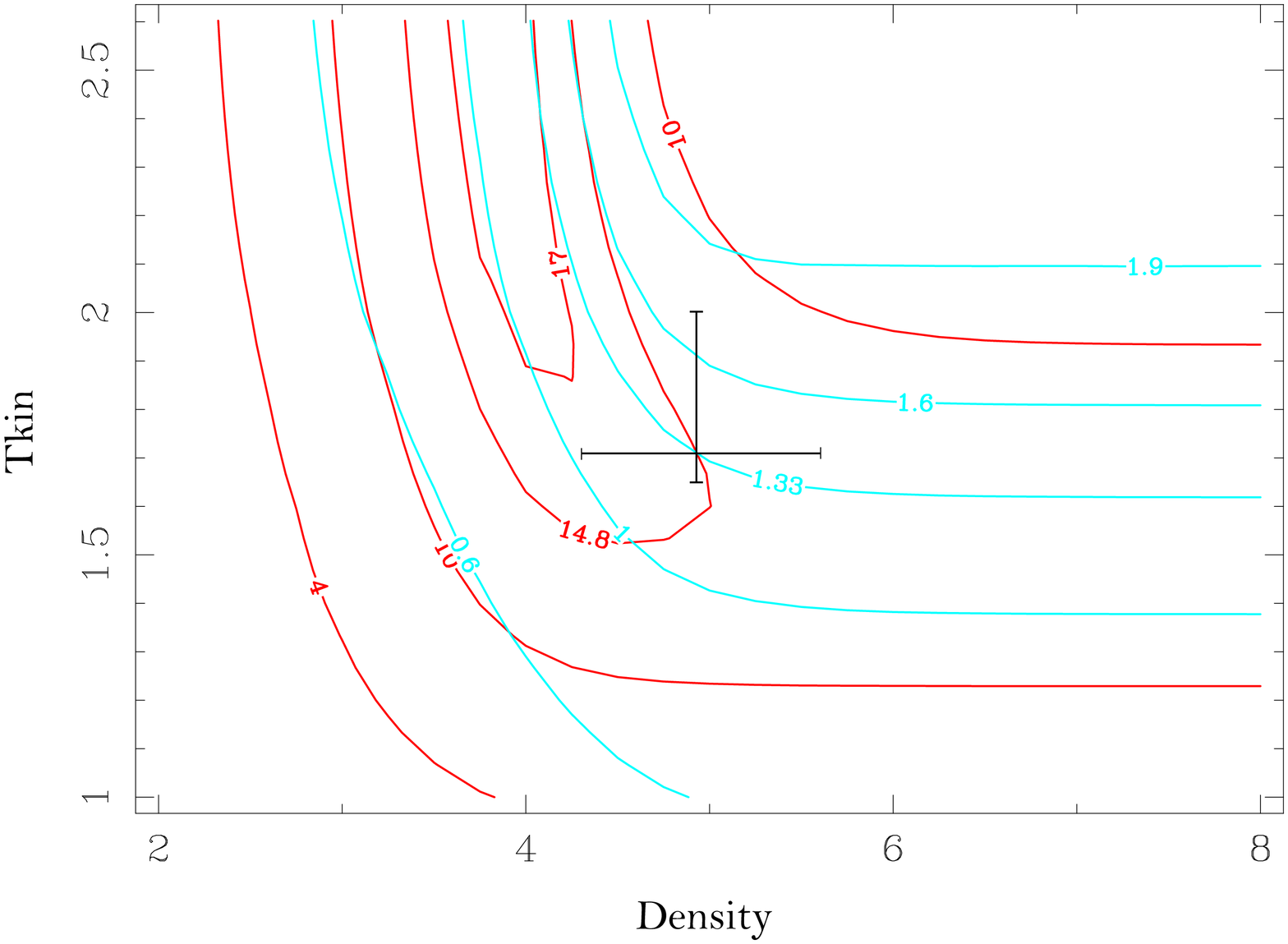}
\caption{LVG model analysis at position A (left panel, $N_{CO}=4.6\times10^{16} cm^{-2}$) and B
(right panel, $N_{CO}=1.2\times10^{16} cm^{-2}$) using the \coTwo and \coThr line. The horizontal and vertical axes
stand for volume density n(H$_{2}$) and kinetic temperature T$_{\rm kin}$
in log scale, respectively. Red curves are line intensity of \coTwo
and blue curves are the line ratios of \coThr/\coTwo. The
results of T$_{\rm kin}$ and n(H$_{2}$) are taken from the cross points of
certain red and blue curves. The error bars indicate the range of the possible solutions listed in Table \ref{Table 2}.
}
   \label{Fig5}
\end{figure}

In order to investigate RDI effects and its physical
characteristics quantitatively, we made further analysis with a large
velocity gradient \citep[LVG,][]{1974Goldreich, 2008Qin} radiative transfer model
 in the MIRIAD software package. In this model, the molecular cloud is
 considered spherically uniform and the non-LTE excitation properties are calculated
 based on the kinetic temperature (T$_{\rm kin}$), the column density per unit
  velocity interval (N$_{\rm CO}$/$\Delta$V), and the H$_{2}$ density (n$_{\rm H_{2}}$).
In this case, we use the line intensity of \coTwo  and the line ratio of \coThr/\coTwo
to determine the volume density n$_{\rm H_{2}}$ and kinetic temperature T$_{\rm kin}$ at the two selected positions
A and B, assuming the column density in a range of $1\times10^{14}$ to $1\times10^{18} cm^{-2}$, and $\Delta$V=1 \kms.
Representative solutions that can be fitted to the observations are listed in Table \ref{Table 2}. A typical case is shown
in Figure \ref{Fig5}.

\begin{table}[h!!!]

\small
\centering

\begin{minipage}[]{120mm}
\caption[]{ Solutions of LVG analysis }
\label{Table 2}\end{minipage}

\tabcolsep 6mm
 \begin{tabular}{clcl}
  \hline\noalign{\smallskip}
Position    &N$_{\rm CO}$ (10$^{16}$ cm$^{-2}$)     &n(H$_{2}$) (10$^{4}$ cm$^{-3}$)    &T$_{\rm kin}$ (K)                 \\
  \hline\noalign{\smallskip}
A  &3.2  &1.6    &40   \\
   &6.8  &3.2    &31   \\
   &10  &4.0    &31   \\
   &22  &2.0    &30   \\
   &46  &1.0    &30   \\
   &68  &0.63    &30   \\
   &100   &0.20    &32   \\
B  &1.1  &4.0    &71   \\
   &1.2  &8.9    &50   \\
   &1.3  &50    &45   \\

  \noalign{\smallskip}\hline
\end{tabular}
\end{table}

For position A, various column densities are possible to explain the observation results self-consistently
 covering a range from $3\times10^{16}$ to $1\times10^{18} cm^{-2}$, with corresponding volume densities ranging from $0.2\times10^4$ to $4.0\times10^4 cm^{-3}$. The average value of the possible volume densities is $2.0\times10^{4} cm^{-3}$.
For position B, the few self-consistent solutions require the column densities in a tiny range of $1.1\times10^{16}$  to $1.3\times10^{16} cm^{-2}$. The average density derived is $1.5\times10^{5} cm^{-3}$, nearly one order of magnitude higher than that at position A.
The average kinematic temperature of B (61 K) is also twice as high as that of A (32 K). Even in the cases that both positions have similar densities (e.g. $n(H_2)=4.0\times10^4 cm^{-3}$), the kinematic temperature required at position B is still 2.3 times than that at position A. Therefore we can conclude that the reversed line ratios along the bright rim are produced
 by either high densities, or high temperatures, and likely both. Both possibilities can be explained by radiation and pressure of the ongoing RDI, where ionizing shock wave of the RDI propagates into the molecular cloud, compressing it at the edge, and thus makes the density and temperature of the edge even higher than the center.

In comparison with the LVG analysis results, we have also calculated the column densities assuming LTE conditions.
Taking the upper energy level temperature of 16.6 K \citep{20081Qin}, the column density of CO is given by \citep{1991Garden}:
\begin{equation}
N_{CO}=1.08\times10^{13}\frac{T_{ex}}{exp(-16.6/T_{ex})} \int T_{mb}dv \,(cm^{-2})
\label{Eq1}
\end{equation}
and
\begin{equation}
T_{ex}=\frac{hv}{k}(ln\{1+\frac{hv}{k}[\frac{T_{mb}}{f}+\frac{hv}{k}(exp(\frac{hv}{kT_{bg}})-1)^{-1}]^{-1}\})^{-1} \,(K)
\label{Eq2}
\end{equation}
where the beam filling factor $f$ is assumed to be 1, and the cosmic background temperature $T_{bg}$ is taken as 2.732 K.
With the \coTwo line properties listed in Table \ref{Table 1}, we derive a CO column density of $4.0\times10^{16} cm^{-2}$
for position A, and $1.9\times10^{16} cm^{-2}$ for position B. Because \coTwo is always considered optically thick,
 the column density derived here should be a lower limit. Compared with the column densities from LVG analysis,
 we find that LVG results at position A roughly agree with the LTE calculation, but for position B, all the LVG solutions
  are lower than the LTE lower limit.
This result agrees with the scenario that at position B the LVG
calculation requires less column density than LTE to explain the
observing results because it takes into account enhanced photon escape
probability due to greater velocity gradient at the bright rim, where
the nearby HII region agitates the molecular cloud with strong FUV
radiation pressure and heating, exhibiting a case of RDI.

\section{Summary}
\label{sect:summary}

We present analysis of large scale mapping observations of the molecular lines
\coTwo \coThr \TcoTwo and  \TcoThr transition emissions toward the Cepheus B molecular
cloud. The target lies at the edge of HII region S155, and is
surrounded by an optically bright rim. The CO $J$=3-2 emission is
stronger than $J$=2-1 emission along the bright rim.
PV diagrams of this region show a velocity gradient from the head of the cloud to the tail, which is characteristic of RDI.
 According to the LVG analysis, the density and temperature at the edge of the cloud are higher than that at the center.
All of our results provide evidences that RDI is taking place in Cepheus B molecular cloud.

\normalem
\begin{acknowledgements}

We would like to thank Sheng-Li Qin, Martin Miller and Nimei Chen
for permission to use observation data and useful directions and discussions. We would also like to
thank Rui Xue for his kind and helpful advice.

\end{acknowledgements}



\label{lastpage}

\end{document}